\documentclass[a4paper,twocolumn]{article}

\usepackage{microtype}
\usepackage{amsmath}
\usepackage{subcaption}
\usepackage{graphicx}
\usepackage[english]{babel}

\usepackage[style=ieee,citestyle=numeric-comp,backend=bibtex]{biblatex}
\bibliography{bibliography.bib}
\newcommand\unit[1]{\,\ensuremath{\mathrm{#1}}}
\newcommand\tmin{\text{min}}
\newcommand\tmax{\text{max}}

\begin{document}

\title{
  Synthesis of Ultra-Wideband TEM Horn with Inhomogeneous Dielectric Medium}

\author{
  G.\,K.\,Uskov,
  P.\,A.\,Kretov,
  V.\,A.\,Stepkin,
  N.\,S.\,Sbitnev,
  and A.\,M.\,Bobreshov
  \thanks{The work was supported by the Russian Federation President's
          grant for young doctors (project~MD-7902.2016.9).}
  \thanks{The authors are with Voronezh State University,
          1,~Universitetskaya~pl., 394018,~Voronezh, Russia.}}
\maketitle

\begin{abstract}
  There were proposed new formulas for the dielectric
  medium permittivity distribution along spatial coordinates in the
  space between a linear TEM-horn's leafs,
  which were obtained according to the rules of geometric optics
  in the assumption of the phase center being lumped or distributed.
  The formulas has been checked by
  FDTD simulation of an ultra-wideband signal excitation;
  two cases were studied---when dielectric medium is present,
  and absent.
  For these both cases, there was obtained voltage steady-wave ratio,
  as well as radiation patterns at a set of frequencies up to 20~GHz;
  the results comparison was performed.
\end{abstract}


\section{Introduction}

These days TEM-horns of various types are widely used
as antennas in ultra-wideband (UWB) radiolocation and communication;
TEM-horns earned their popularity for having wide range of operational
frequencies, while being extremely easy for manufacturing.
Unfortunately, these antennas have a number of drawbacks too;
for example, they must have relatively large electric size
to radiate signals efficiently
and may have low-gain frequency gaps.
These drawbacks can be partially worked around by carefully choosing
antenna size and shape\,\cite{
  Chung2005,Bobreshov2012,Bobreshov2013,Chen2013,Mehrdadian2014};
moreover, it is possible to make horn operational frequency range
even wider,
usually at the cost of further antenna size
increasing\,\cite{Lee2004}.

There is another approach for improving TEM-horns' gain and
patterns which is based on combining antennas with dielectric objects.
These objects usually are complex-shaped lens manufactured from a solid
material which are placed at aperture plane
or beyond it\,\cite{Holzman2004,Efimova2012};
while improving gain and patterns,
these techniques have a common downside of increasing energy
reflection back to feeding line and, hence, increasing voltage
steady-wave ratio (VSWR) at higher frequencies.

\begin{figure}[!b]
  \centering
  \includegraphics[width=1.1\linewidth]{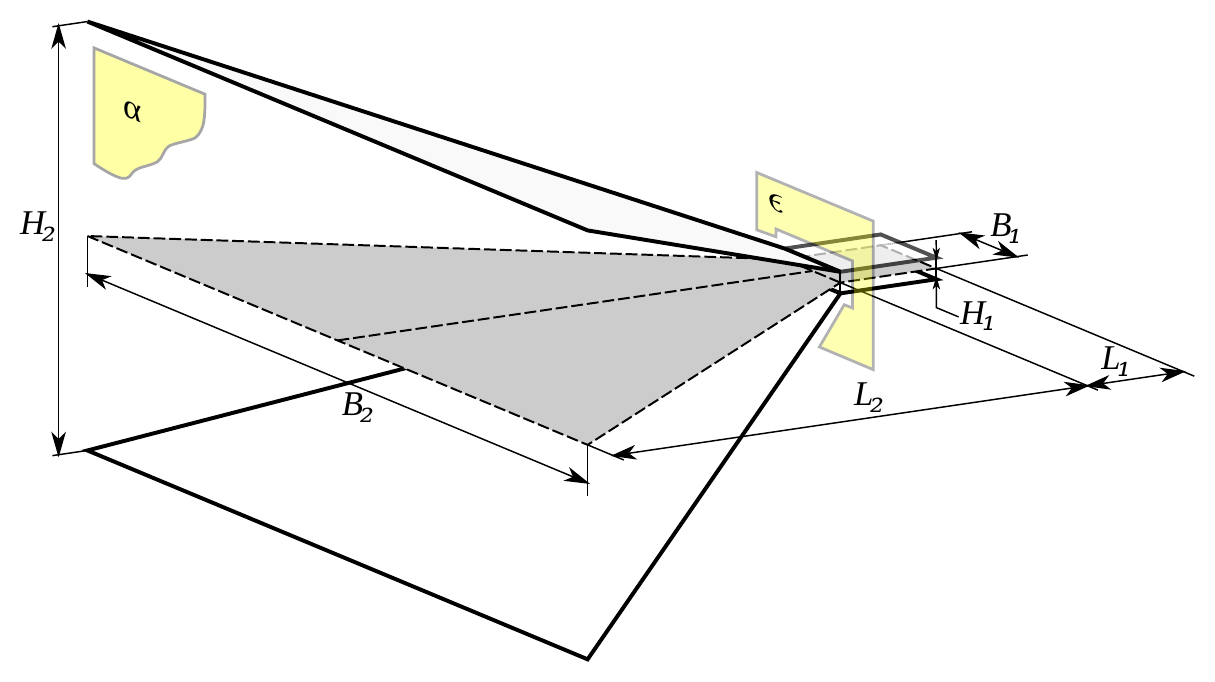}
  \caption{
    TEM horn antenna of the interest, with size markers shown.
    Here~$\alpha$ is the aperture plane,
         $\epsilon$~is the excitation plane.}
  \label{fig:horn}
\end{figure}

However, there is way to improve horn's directivity
without sacrificing VSWR level.
Its essence is filling the space between horn leafs by
dielectric medium with spatially varying permittivity;
manufacturing of such filling objects was
a major technological problem in near past,
which can be successfully solved in present time
using a number of techniques, including, but not limited to,
3D~printing\,\cite{Liang2014,Molina2016}.
For example, Molina and Hasselbarth\,\cite{Molina2016}
accompanied a planar slot radiator by a dielectric lense
created by layer-by-layer backing a mixture of alumina ceramic
powder and microscopic hollow glass spheres.

In this work, for a given TEM horn with flat trapezoidal leafs
(see~Fig.\,\ref{fig:horn} for its schematic view) we
(a)~derive dielectric filling permittivity distribution from
    geometric optics model of radiation process, and
(b)~use time-domain computer simulation of ultra-short pulse
    propagation to check whether adding this filling actually
    improves horn characteristics.

\section{Wavefront planarization}

Let's assume for the rest of the paper that
UWB signals propagate through the space like they are beams of light.
Obviously, this assumption is quite strong for
frequency order of tens gigahertz,
but it lets us derive simple, but useful material distributions.
This approach is not uncommon in antenna
design\,\cite{Karttunen2013,Aghanejad2012}, either.

In the following subsections we present
two similarly obtained distribution formulas.
For the first one, we assumed that the whole electromagnetic wave
was emitted from the single point
(we call it a \emph{lumped phase center} in this paper);
for the second one, we uniformly stretched that phase center
across horn's excitation plane to form a \emph{distributed phase center}.

\begin{figure}[!t]
  \centering
  \begin{subfigure}[b]{0.49\linewidth}
    \centering
    \includegraphics[width=\linewidth]{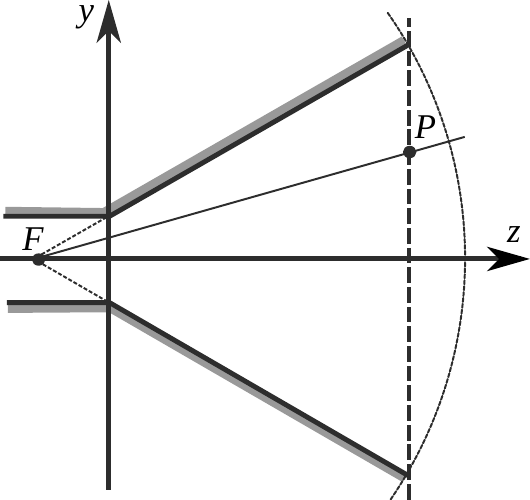}
    \caption{lumped phase center}
    \label{fig:planarization:lumped}
  \end{subfigure}
%
  \begin{subfigure}[b]{0.49\linewidth}
    \centering
    \includegraphics[width=\linewidth]{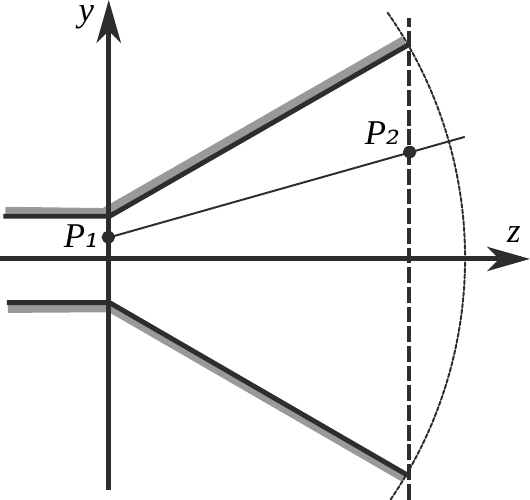}
    \caption{distributed phase center}
    \label{fig:planarization:distributed}
  \end{subfigure}

  \caption{To the wavefront planarization procedures.}
  \label{fig:planarization}
\end{figure}

\begin{figure*}[!t]
  \centering
  \begin{subfigure}[b]{0.32\linewidth}
    \centering
    \includegraphics[width=\linewidth]{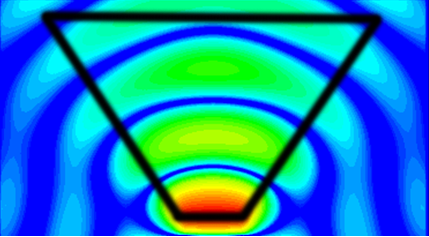}
    \caption{no filling}
    \label{fig:field_image_horn=none}
  \end{subfigure}
  \begin{subfigure}[b]{0.32\linewidth}
    \centering
    \includegraphics[width=\linewidth]{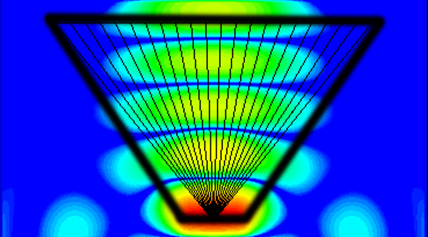}
    \caption{lumped phase center}
    \label{fig:field_image_horn=orig}
  \end{subfigure}
  \begin{subfigure}[b]{0.32\linewidth}
    \centering
    \includegraphics[width=\linewidth]{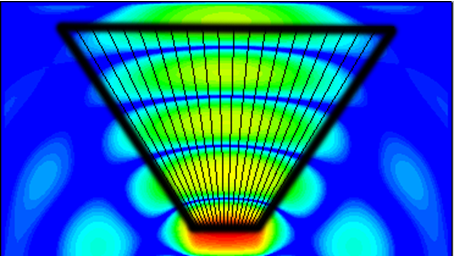}
    \caption{distributed phase center}
    \label{fig:field_image_horn=mod}
  \end{subfigure}
  \caption{
    Electric field magnitude in horn's horizontal symmetry
    plane at~15\unit{GHz} (time-domain simulation).}
  \label{fig:field_image}
\end{figure*}

\subsection{Lumped phase center}

Let's choose a system of Cartesian coordinates such that~$x$ axis is
horizontal, $y$~axis is vertical, and~$z$ is TEM-horn's main direction
(as shown in~Fig.\,\ref{fig:planarization:lumped}); and put the origin
at the point where the horn's symmetry axes intersect
the excitation plane.
Also consider the point~$F$ which is the midpoint of the line
formed by horn leafs' continuations intersection;
let this~$F$ be the lumped phase center.

If electromagnetic beams are simultaneously emitted from~$F$, then
after a certain amount of time they form a spherical wavefront.
The idea is to slow down central beams by increasing dielectric
permittivity along their trajectories, and, hence, turn the spherical
wavefront into a plane.

Let~$R_\tmax$ be the longest possible straight path from~$F$ to the
aperture rectangle as $R_\tmax$, which is the line pointing to
a leaf's trapeze apex.
Since that path is the longest, the material along it must have
the lowest dielectric permittivity among all beam paths;
call it~$\varepsilon_\tmin$.
It is safe to assume that $\varepsilon_\tmin = 1$, hence
\begin{equation*}
    t_\tmax = \frac{R_\tmax}{c}
\end{equation*}
is the time required for a beam to travel along the~$R_\tmax$ path,
where~$c$ is the speed of light in free space.

Let's now consider a beam propagating in
arbitrary direction inside the horn.
Let~$R$ be the distance between~$F$ and the point~$P$
where the beam intersects the aperture plane~$\alpha$,
so the time required for the beam to travel to that point is
\begin{equation*}
    t = \frac{R}{v},
\end{equation*}
where~$v$ is the speed of electromagnetic wave in the dielectric
material located along the path.

For the $R_\tmax$-beam and $R$-beam to reach the horn exit plane
simultaneously, the following condition must be met:
\begin{equation*}
    \frac{R_\tmax}{c} = \frac{R}{v}.
\end{equation*}
Taking into account that $\varepsilon = (c/v)^2$, the following
expression for the dielectric permittivity inside the horn can be
obtained:
\begin{equation}
    \label{eq:lumped:epsilon}
    \varepsilon = \frac{R^2_\tmax}{R^2}.
\end{equation}
In the coordinate system we introduced above, it can be stated that
\begin{align*}
    R^2_\tmax &= \frac14 \left( B_2^2 + H_2^2 \right)
                      + \left( L_2 + l \right)^2, \\
    R^2      &= x^2 + y^2 + \left( L_2 + l \right)^2,
\end{align*}
where~$l$ is the distance between~$F$ and the excitation plane,
$x$ and $y$ are the coordinates of a point~$P$,
$|x| \leq B_2/2$, $|y| \leq H_2/2$.
If we substitute the above expression into
equation~(\ref{eq:lumped:epsilon}),
we get the following expression:
\begin{equation}
  \label{eq:lumped:distribution}
    \varepsilon = \frac
        {\frac14 \left( B_2^2 + H_2^2 \right) + \left( L_2 + l \right)^2}
        {x^2 + y^2 + \left( L_2 + l \right)^2}.
\end{equation}

Equation~(\ref{eq:lumped:distribution}) gives us a formula of
material distribution along
straight lines connecting~$F$ and each point at the aperture plane.
It can be also rewritten as the
$\varepsilon(\theta) \propto \cos^2 \theta$
dependence,
where~$\theta$ is the angle between the beam and the
horn's main direction.

\subsection{Distributed phase center}

Despite the above section's approach has proven itself to work in
practice, its propagation model is simplistic and cannot reflect
reality very well. Let's elaborate the model to make it more
adequate.

Computer simulation shows (Fig.\,\ref{fig:field_image_horn=none})
that TEM horn does not emit waves from just a single point;
instead, the irradiation process involves the whole
excitation plane.
This means that the phase center should not be though as a lumped
entity, but as a continuum of phase centers distributed on the
excitation plane between horn leafs.
So, let's modify the formula~(\ref{eq:lumped:distribution}) to
convert lumped phase center into distributed.

As in the previous case, we continue to think about the wavefront as a
set of beams, but now these beams are not required to have a common
origin (former point~$F$).
Instead, let each beam start from the point~$P_1=(x_1, y_1)$
at the excitation plane
and intersect the aperture plane at the point $P_2=(x_2, y_2)$.
Let's additionally require the following restriction to satisfy:
\begin{equation}
    \label{eq:nonlumped:restriction}
    \frac{x_1}{B_1} = \frac{x_2}{B_2}, \quad
    \frac{y_1}{H_1} = \frac{y_2}{H_2}.
\end{equation}

In these new circumstances, $R$ and~$R_\tmax$ values can be obtained as
\begin{align*}
  R^2_\tmax &= \frac14 \left[ \left( B_2 - B_1 \right)^2 +
                             \left( H_2 - H_1 \right)^2 \right] +
                             \left( L_2 - L_1 \right)^2,\\
  R^2 &= \left( x_2 - x_1 \right)^2 +
         \left( y_2 - y_1 \right)^2 +
         \left( L_2 - L_1 \right)^2.
\end{align*}
If we substitute the above expressions into
equation~(\ref{eq:lumped:epsilon}) taking into account that
$x_1=\frac{B_1}{B_2} x_2$ and
$y_1=\frac{H_1}{H_2} y_2$,
when the final formula will be
\begin{equation}
  \varepsilon = \frac
    {\frac14 \left[
        \left( B_2 - B_1 \right)^2 +
        \left( H_2 - H_1 \right)^2
     \right]
      + \left( L_2 - L_1 \right)^2}
    {\left(1 - \frac{B_1}{B_2} \right)^2 x_2^2 +
     \left(1 - \frac{H_1}{H_2} \right)^2 y_2^2 +
     \left( L_2 - L_1 \right)^2}.
\end{equation}

\section{Computer simulation model}

In order to check the proposed material distributions, we made their
digital models, suitable for finite-difference
time-domain~(FDTD)\,\cite{Taflove1995} simulation.
This model included a simple TEM horn (Fig.\,\ref{fig:horn}) with
a strip feeding line;
the dimensions were chosen as follows:
\begin{equation*}
\begin{array}{ll}
  L_1 = 30\unit{mm}, & L_2 = 30\unit{mm} \\
  B_1 = 10\unit{mm}, & B_2 = 50\unit{mm} \\
  H_1 = 2\unit{mm},  & H_2 = 50\unit{mm} \\
\end{array}
\end{equation*}

In this model, all antenna and feeder surfaces were approximated by
thin perfect electric conductors.
The whole problem's geometry was fit into
$68\unit{mm}\times 57\unit{mm}\times 63\unit{mm}$
computational domain and split into
$151 \times 135 \times 137$
cuboid Yee cells.
To prevent radiated signal from reflecting back into computational
domain, four-layer convolutional perfectly matched layer (PML)
boundary condition\,\cite{Berenger1996} was applied to domain
edges (each having 5\unit{mm} as the minimum distance to the nearest
conducting surface).

\section{Obtained results}

After simulation run, we obtained horizontal-plane antenna patterns
at 5, 10, 15, and 20\unit{GHz} frequencies,
as well as the feeding line S-parameters.
Additionally, images of electric field were constructed;
they were already presented in Fig.\,\ref{fig:field_image}

Figures~\ref{fig:field_image_horn=orig}
    and~\ref{fig:field_image_horn=mod}
demonstrate that both lumped and distributed phase center approaches
make wavefront a flatter shape
(compared to normal medium, Fig.\,\ref{fig:field_image_horn=none}),
though, not a perfect plane.
However, in the case of the distributed phase center,
more of peripheral aperture is involved in signal formation,
thus increasing antenna's effective area.

\begin{figure*}[p]
  \centering
  \begin{subfigure}[b]{0.48\linewidth}
    \centering
    \includegraphics[width=65mm]{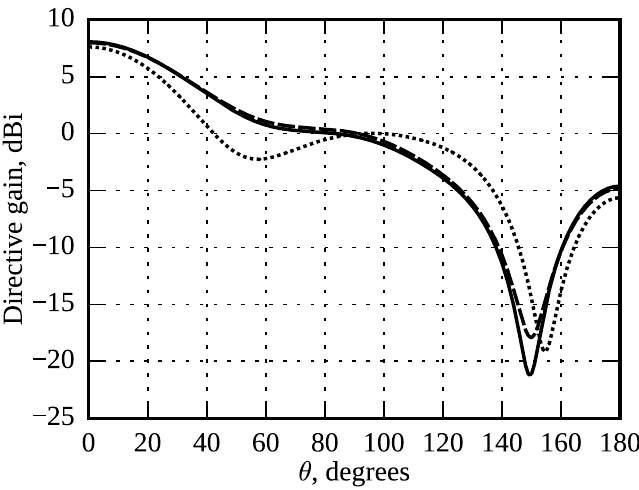}
    \caption{$f = 5\unit{GHz}$}
  \end{subfigure}
  \begin{subfigure}[b]{0.48\linewidth}
    \centering
    \includegraphics[width=65mm]{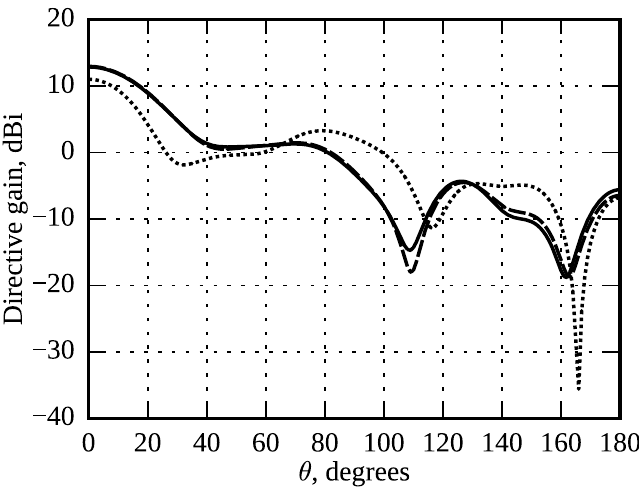}
    \caption{$f = 10\unit{GHz}$}
  \end{subfigure}
  \begin{subfigure}[b]{0.48\linewidth}
    \centering
    \vspace{4mm}
    \includegraphics[width=65mm]{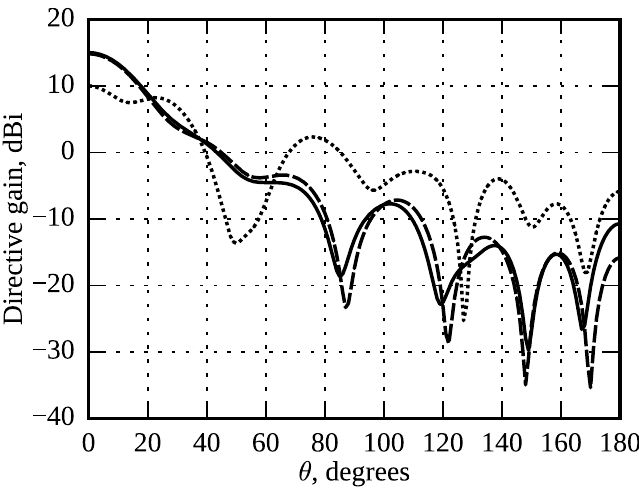}
    \caption{$f = 15\unit{GHz}$}
  \end{subfigure}
  \begin{subfigure}[b]{0.48\linewidth}
    \centering
    \vspace{4mm}
    \includegraphics[width=65mm]{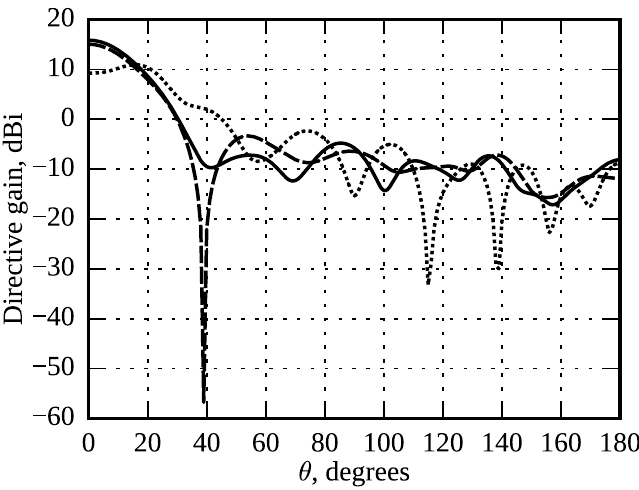}
    \caption{$f = 20\unit{GHz}$}
  \end{subfigure}

  \caption{
    Antenna patterns in the horizontal symmentry plane,
    where~$\theta$ is the angle betwen the direction and $z$~axis.}
  \label{fig:patterns}
\end{figure*}

\begin{figure*}[p]
  \centering
  \begin{subfigure}[b]{0.48\linewidth}
    \centering
    \vspace{4mm}
    \includegraphics[width=65mm]{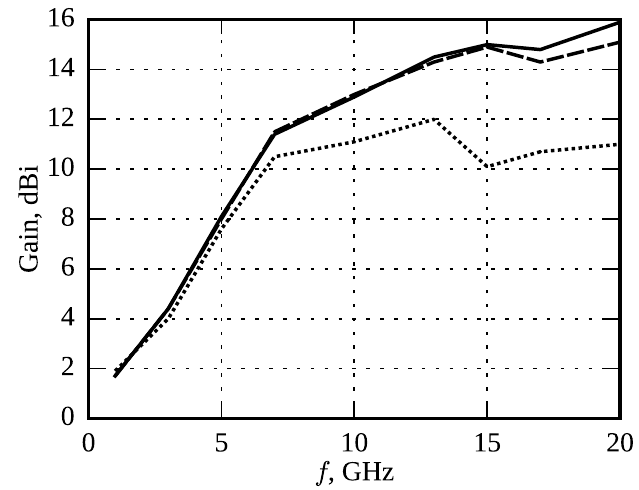}
    \caption{gain}
    \label{fig:gain}
  \end{subfigure}
  \begin{subfigure}[b]{0.48\linewidth}
    \centering
    \vspace{4mm}
    \includegraphics[width=65mm]{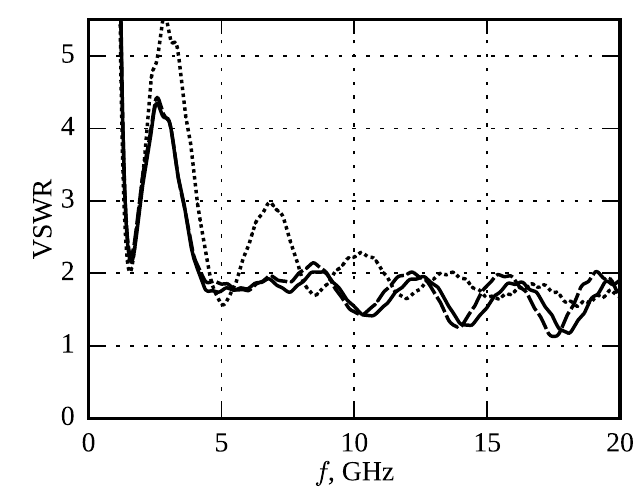}
    \caption{VSWR}
    \label{fig:vswr}
  \end{subfigure}

  \caption{TEM horn frequency parameters.}
\end{figure*}

Obtained patterns are presented in Fig.\,\ref{fig:patterns}.
It can be seen that the presence of dielectric filling
(with either lumped or distributed phase center)
improves directional selectivity of antenna;
more specifically, filled antennas have higher gains at all considered
frequencies~(Fig.\,\ref{fig:gain}), as well as narrower main lobes.
This improvement is more significant for the cases of 15 and 20\unit{GHz}
than for 5 and 10\unit{GHz}.
Moreover, adding dielectric filling prevented main lobe from splitting
at 20\unit{GHz}, which we consider a notable practical result.

Figure~\ref{fig:vswr} shows VSWR frequency dependence
which was calculated from S-parameters obtained after the simulation run.
It shows that the above said gain and directivity improvements did not
compromised antenna matching.

\section{Conclusion}

This paper presented two similar, but different formulas for filling
TEM-horns by inhomogeneous dielectric medium.
They were obtained in the assumption of light-like propagation of
the UWB signal.
Though this assumption is quite unrealistic,
computer simulation confirmed that
employing these formulas for generating dielectric filling
significantly increases horn's gain and directional selectivity
without impairing antenna matching.

Despite the antenna being studied was linear TEM-horn,
the same approach can potentially be applied to more
complex-shaped horns by, for example,
subdividing this complex shape into a number of short linear intervals.

\nocite{*}
\printbibliography

\end{document}